\documentclass[fleqn,twoside]{article}
\usepackage{espcrc2}

\usepackage{graphicx}

\newcommand{\AmS}{{\protect\the\textfont2
  A\kern-.1667em\lower.5ex\hbox{M}\kern-.125emS}}
\def\0{\over } \def\2{{1\over2}} \def\4{{1\over4}}
\def\5{\hat } \def\6{\partial }

   \def\d{\delta }
\def\e{\epsilon }

 \def\o{\omega }
\def\ph{\varphi }

\def\({\left(} \def\){\right)} \def\<{\langle } \def\>{\rangle }

\newcommand{\bea}{\begin{eqnarray}}
\newcommand{\eea}{\end{eqnarray}}
\newcommand{\be}{\begin{equation}}
\newcommand{\ee}{\end{equation}}
\newcommand{\nn}{\nonumber\\ }
\newcommand \beq{\begin{eqnarray}}
\newcommand \eeq{\end{eqnarray}}

\def\Im{{\,\mathrm{Im}\,}}
\def\Re{{\,\mathrm{Re}\,}}
\def\tr{{\,\mathrm{tr}\,}}

\title{HTL quasiparticle picture of the thermodynamics of QCD$^\star$}

\author{A. Rebhan\address{Institut f\"ur Theoretische Physik,
Technische Universit\"at Wien\\
Wiedner Hauptstr. 8--10, A-1040 Vienna, Austria
}}
       
\begin{document}

\begin{abstract}
Starting from a nonperturbative expression for entropy and density
obtained from $\Phi$-derivable two-loop approximations to the
thermodynamic potential, a quasiparticle model for the
thermodynamics of QCD can be developed which incorporates
the physics of hard thermal loops and leads to a reorganization
of the otherwise ill-behaved thermal perturbation theory through
order $\alpha_s^{3/2}$. Some details of this reorganization are
discussed and the differences to simpler quasiparticle
models highlighted. A comparison with available lattice
data shows remarkable agreement down to temperatures of
$\sim 2.5 T_c$.
\vspace{1pc}
\end{abstract}

\maketitle

\begin{picture}(0,0)(0,0)
\put(-10,270){\large\tt TUW 01-23}
\put(-10,-380){\rule{60pt}{0.2pt}}
\put(-10,-400){$^\star$Talk given at the International Conference
on Statistical QCD, Bielefeld, Germany, August 26--30, 2001}
\end{picture}

\section{Entropy and density in $\Phi$-derivable approximations
}

In the so-called $\Phi$-derivable approximations \cite{Baym:1962}
obtained by truncating the skeleton functional ($\Phi$)
in the Luttinger-Ward representation of the thermodynamic
potential \cite{Luttinger:1960},
the expression for the entropy density takes a remarkable form:
for example in a scalar theory $g\ph^3+g^2\ph^4$
it can be written as
\bea\label{Ssc}
{\cal S}=-\int\!\!{d^4k\0(2\pi)^4}{\6n_{\rm BE}(\o)\0\6T} \Im \log D^{-1}(\o,k) \nn
+\int\!\!{d^4k\0(2\pi)^4}{\6n(\o)\0\6T} \Im\Pi(\o,k) \Re D(\o,k)+{\cal S}'
\eea
where $D$ and $\Pi$ are self-consistent propagator and self-energy, resp.,
and one finds
\be\label{SP0}
{\cal S}'=O(\mbox{3-loop})=O(g^4).
\ee
The fact that ${\cal S}'=0$ in 2-loop $\Phi$-derivable approximations,
which
has been observed first in Fermi-liquid theory by Riedel \cite{Riedel:1968},
turns out to hold rather generally \cite{Vanderheyden:1998ph,Blaizot:1999ip,Blaizot:1999ap,Blaizot:2000fc,Blaizot:2001BI}. In particular, the
2-loop contribution of Dirac fermions to the entropy
density reads
\bea\label{SQCDF}
{\cal S}_f = -2\int\!\!{d^4k\0(2\pi)^4}{\6n_{\rm FD}(\o)\0\6T}\,
\tr\biggl\{\Im\log(\gamma_0 S^{-1}) \nonumber\\
\,-\,\Im(\gamma_0 \Sigma)
\Re(S\gamma_0)\biggr\},\eea
For nonvanishing chemical potentials $\mu_i$, this can be
extended to the fermion number
densities ${\cal N}_i$, which are obtained by replacing the explicit
derivative $\6/\6 T$ in (\ref{SQCDF}) by $\6/\6\mu_i$, with
\be
{\cal S}'_f=O(\mbox{3-loop})={\cal N}'_i
\ee

Explicit interactions at 2-loop order are thus completely
encoded in the dressed (quasiparticle) propagators $D$, $S$, $\ldots$,
which makes entropy and density a preferred starting point
for a quasiparticle description of thermodynamic quantities---in
fact any thermodynamic quantity, since
apart from a single integration constant (which
in QCD corresponds to the inherently nonperturbative bag constant)
the grand canonical potential can be reconstructed from
entropy and density.

If a quasiparticle picture in the sense of Landau is applicable,
that is, if a large fraction of the possibly strong
interactions can be accounted for by `propagator renormalization',
the above dressed 2-loop expressions should provide
a reasonable approximation to thermodynamic quantities
and the
residual quasiparticle interactions $\sim g^4$
described by ${\cal S}'$ or
${\cal N}'$ should be comparatively weak.

Technically, an important point turns out to be that
the above expressions for entropy and density are
manifestly ultraviolet finite, as all
integrals involve ${\6n/\6T}$ or ${\6n/\6\mu}$ which
go to zero exponentially for {\em both} positive and negative
frequencies. In relativistic field theories, this is
a great advantage, because it allows one to use these functionals
in a nonperturbative manner once finite approximations for
the propagators and self-energies have been obtained, whereas usually
the renormalization programme requires systematic expansions
(and truncations) in powers of the coupling constant.\footnote{In
the recent work of Ref.~\cite{vanHees:2001ik} the renormalizability
of $\Phi$-derivable approximations has been postulated, which is
however possibly in contradiction with the explicit calculations
of Ref.~\cite{Braaten:2001vr}.}

At finite temperature, truncated perturbative series turn out
to have extremely poor apparent convergence behaviour
\cite{Drummond:1997cw} and this problem arises with the
appearance of odd powers in the coupling such as
$g^3$ (the so-called plasmon effect\footnote{Caused in fact by
the appearance of screening (Debye) masses as opposed to
the dynamical plasmon mass.}). In QCD this plasmon effect
spoils apparent convergence up to extremely high temperatures
$\gg 10^5 T_c$. However, this breakdown of perturbation theory
does not necessarily have anything to do with the nonabelian character
of QCD, as a similar phenomenon occurs in virtually any
field theory for coupling constants where $T=0$ perturbation
theory still looks applicable.

On the other hand, in expressions such as (\ref{Ssc}) and (\ref{SQCDF})
the coupling constants are all contained in dressed propagators and
self-energies, which themselves are buried in logarithmic or
fractional expressions. If there are contributions which when
expanded out in powers of the coupling have uncomfortably
large coefficients, they do not necessarily lead to large
corrections as they are kept together with higher-order terms
(which usually would be discarded) such that final results
behave more like logarithms or fractions rather than
polynomials.

Starting from the 2-loop $\Phi$-derivable entropy (density),
one should therefore be in a much more favourable position
to investigate whether and to what extent
a quasiparticle picture such as that
underlying hard-thermal-loop (HTL) 
perturbation theory \cite{Braaten:1990mz} is applicable.

In the following I shall describe the results obtained
by employing HTL propagators and self-energies and
certain corrections thereof in the above
nonperturbative expressions for entropy (and density).

This will only provide an approximation to the
self-consistent 1-loop gap equation that pertains
to the $\Phi$-derivable 2-loop approximation.
A complete solution of the self-consistent gap equation
is presumably prohibitively complicated, but apart from
that, not really desirable:
While eqs. (\ref{Ssc}) and (\ref{SQCDF}) are
UV finite, the associated 1-loop gap equations are not.
Moreover, in gauge theories these equations are gauge
dependent, as $\Phi$-derivable approximations do not
respect local gauge invariance. The approximately
self-consistent solutions that will be discussed below
evade both problems by providing finite, gauge invariant
and gauge fixing independent approximations, and they
do so by dropping terms that are anyway beyond the
accuracy of any 2-loop approximation, i.e., of
order $g^4$ or higher.

\section{Approximate propagators}

In thermal perturbation theory one has to distinguish
hard and soft momentum scales. At high temperature (in
QCD: sufficiently above the deconfinement transition),
the dominant contribution to the grand canonical potential
comes from elementary quanta with energies and momenta
$\sim T$, which will be called ``hard''. Collective
phenomena arise at the scale $gT$, which will be called
``soft''. Even when $g \sim 1$, as will be the case
of interest in QCD, this distinction may still make
sense, if the scale of hard physics is actually $2\pi T$
(the spacing of the Matsubara frequencies)rather than just $T$, as
frequently argued \cite{Braaten:1995cm}.

At {\em soft} momentum scales, the leading terms in the
self-energies are given by the HTL self-energies, which
for gauge bosons and (massless) fermions contain
two separate structure functions each and read
\cite{LeB:TFT}
\bea
\label{PiL}
\5\Pi_L(\o,k)=\5m_D^2\left[1-{\o\02k}\log{\o+k\0\o-k}\right],\\
\label{PiT}
\5\Pi_T(\o,k)=\frac{1}{2}\left[\5m_D^2+\,\frac{\o^2 - k^2}{k^2}\,
\5\Pi_L\right],\\
\label{SIGHTL}
\hat\Sigma_\pm(\omega,k)={\hat M^2\0k}\,\left(1\,-\,
\frac{\omega\mp k}{2k}\,\log\,\frac{\omega + k}{\omega - k}
\right),\eea
with
\bea
\hat m_D^2=(N+{N_f\02}){g^2T^2\03}+\sum_f{g^2\mu_f^2\02\pi^2},\\
\hat M_f^2={g^2C_F\08}(T^2+{\mu_f^2\0\pi^2}).
\eea

The gauge boson propagator involves two different
propagators, a spatially transverse one,
\be
\hat D_T(\o,k)=[-\o^2+k^2+\5\Pi_T(\o,k)]^{-1},
\ee
and one that only appears in the thermal medium,
\be
\hat D_L(\o,k)=-[k^2+\5\Pi_L(\o,k)]^{-1}.
\ee
Similarly, the fermion propagator has a ``normal'' branch
$S_+\propto [\o-(k+\Sigma_+)]^{-1}$, and one that is
a purely collective effect, $S_-\propto [\o+(k+\Sigma_-)]^{-1}$,
where the label $(-)$ refers to the fact that
chirality and helicity have opposite signs.

These HTL propagators have poles at real frequencies $\o$ for all $k$,
corresponding to (undamped) quasiparticles
with momentum-dependent thermal masses, with the additional
branches disappearing from the spectrum for $k/gT \gg 1$
because of exponentially vanishing residues.
The normal branches on the other hand, approach constant
asymptotic masses $m_\infty^2=\2 \hat m_D^2$ and $M_\infty^2=2 \hat M^2$,
resp., in this limit.

At {\em hard} momenta, the above HTL expressions are actually
invalid, except for $|\o^2-k^2| \ll T^2$. This is however
the region where all the poles of the HTL propagators
lie, so that the latter still give the correct dispersion law
of quasiparticles at
leading order, with next-to-leading order
corrections calculable within standard HTL perturbation theory.



\subsection{Leading-order terms in entropy and
density}

Let us now inspect how the leading-order interaction terms
of the grand canonical potential arise in the self-consistent
entropy (and density) functionals (\ref{Ssc}), (\ref{SQCDF}),
which at two-loop order
are determined by interaction-free quasiparticle propagators.

Since $\Pi\propto g^2$, these must be reproduced by linearizing
entropy and density in $\Pi$. In the pure glue case, the
relevant terms in the entropy read
\bea\label{SG2}
{\cal S}_2 =-2N_g\int\!\!{d^4k\0(2\pi)^4}\,{\6n\0\6T}\biggl\{-
\Im\frac{\Pi_T^{(2)}}{\o^2-k^2}\nonumber\\+\Im{\Pi_T^{(2)}}\Re\frac
{1}{\o^2-k^2}\biggr\}\nonumber\\
 =2N_g\int\!\!{d^4k\0(2\pi)^4}\,{\6n\0\6T}\,\Re{\Pi_T^{(2)}}\!\!\!
\underbrace{\Im\frac{1}{\o^2-k^2}}_{-\pi\e(\o)\d(\o^2-k^2)},\eea
where only the $g^2$ part of the transverse polarization tensor,
$\Pi_T^{(2)}$ is needed. The latter is seen to contribute only
its light-cone value 
\be\label{minfty}
\Pi_T^{(2)}(\o^2=k^2)=g^2NT^2/6 \equiv m_\infty^2,
\ee
yielding ${\cal S}_2 =-N_g  m_\infty^2 T/6$ with $N_g=N^2-1$.

Analogous results are obtained for the fermionic contributions,
and also in the case of fermion densities. The order-$g^2$
terms are simply and generally given by the (leading-order) thermal masses
of hard particles according to
\beq\label{SNBF2}
&&{\cal S}_2 =-  T\left\{\sum_B { m_{\infty\,B}^2 \0 12}\,+\,
\sum_F { M_{\infty\,F}^2\0 24}\right\},\\
&&{\cal N}_2=-\,{1\0 8\pi^2}\sum_F \mu_F M_{\infty\,F}^2,\eea
where the sums run over all the bosonic ($B$) and fermionic ($F$)
degrees of freedom (explicitly counting spin degrees of freedom).

\subsection{Plasmon terms}

In conventionally resummed perturbation theory, the so-called
plasmon term $\propto g^3$ in the thermodynamic potential
arises from the appearance of the Debye mass in the
electrostatic propagator. Only the Matsubara zero modes
are able to contribute an odd power of the Debye mass (and
thus an odd power in the coupling) through ring resummation
according to
\bea\label{P31}
\lefteqn{P_3=-N_gT\int\!\!{d^3 k\0(2\pi)^3}\,\left[\log\left(
1+\frac{\5 m_D^2}{k^2}\right)-\frac{\5 m_D^2}{k^2}\right]}\nonumber\\
&&=N_g \frac{\hat m_D^3 T}{12\pi}.\eea

In the self-consistent entropy expression, this plasmon term
arises rather differently. It no longer suffices to resum
the Debye mass for zero modes. Instead, there are soft as
well as hard contributions to order $g^3$.

At soft momenta, these contributions come from HTL propagators
and self-energies in the form (for pure glue and with
implicit sums over colour and polarization states)
\bea\label{SHTLD}
\lefteqn{{\cal S}^{\rm soft}_3=
-\int\!\!{d^4k\0(2\pi)^4}\,\frac{1}{\omega}\biggl\{\Im
\Bigl[\log(1+ D_0\hat\Pi) - \hat\Pi D_0\Bigr]}\nonumber\\ 
&&\qquad\qquad\qquad\qquad-\Im\hat \Pi\Re(\hat D-D_0)\biggr\}\nonumber\\
&&={\6 P_3\0\6 T}\Big|_{\hat m_D}
+\Delta {\cal S}_3,\eeq
with
\bea\label{DELTASS}
\lefteqn{\Delta {\cal S}_3\equiv 
N_g\int\!\!{d^4k\0(2\pi)^4}\,{1\0\o}\,
\biggl\{2\Im\5\Pi_T \Re \Bigl(\5 D_T - D_T^{(0)}\Bigr)}\nonumber\\
&&\qquad\qquad\qquad\quad
-\Im\5\Pi_L \Re\Bigl(\5 D_L-D_L^{(0)}\Bigr)\biggr\}\nn
&&\equiv \Delta {\cal S}_T^{(3)}+\Delta {\cal S}_L^{(3)}\,.\eea
We found numerically that $\Delta {\cal S}_3=0$ 
by cancellations in more than
8 significant digits, without being able to gain more
fundamental insight into the peculiar sum rule $\Delta {\cal S}_T^{(3)}
=-\Delta {\cal S}_L^{(3)}$, which emerges only after carrying
out both, the frequency and the momentum integrations in (\ref{DELTASS}).
In the pure-glue case we thus have 
\be{\cal S}^{\rm soft}_{3}={\6 P_3\0\6 T}\Big|_{\hat m_D}=\frac14 {\cal S}_3.
\ee

At hard momenta, the only possibility for contributions $\sim g^3$ is
through NLO corrections to the spectral properties of
the hard excitations. Indeed, one can show \cite{Blaizot:2000fc} 
that the remaining three quarters of the plasmon effect
are provided by
\be\label{ST32}
{\cal S}_3^{\rm hard}=
- N_g\!\int\!\!{d^3k\0(2\pi)^3}{1\0k}{\6n(k)\0\6T}
\Re\delta \Pi_T(\o\!=\!k).\;\ee
$\Re\delta \Pi_T(\o\!=\!k)$ is a momentum-dependent correction
$\sim g^3T^2$
to the asymptotic thermal mass of transverse gluons, and it
is calculable from standard HTL perturbation theory without
being afflicted by the IR sensitivity that arises in
the imaginary part \cite{Flechsig:1995sk}.

Clearly, the self-consistent entropy involves a ``massive'' reorganization
of thermal perturbation theory. Instead of coming exclusively from
a modification of the static propagators, the plasmon effect
is now spread over the entire spectral properties of quasiparticles
as given by HTL propagators at soft momenta and HTL masses plus
NLO corrections at hard momenta. While this seems to be just
an extravagant complication from the point of view of perturbation
theory, with regard to the self-consistent entropy
it is gratifying that the plasmon effect is
encoded in the spectral data of (HTL) quasiparticles
in a more uniform manner, with the bulk of the plasmon effect
coming in fact from the dominant hard degrees of freedom.

All this can be extended in a straightforward manner to $N_f\not=0$.
In the case of finite chemical potential but high enough
temperature $T\gg \hat m_D$, one finds that more than 1/4 of ${\cal S}_3$
is coming from the soft momentum region, whereas all of the
plasmon term in the density, ${\cal N}_3$, is due
to NLO corrections of the asymptotic thermal fermion masses.

\section{Nonperturbative usage}

As mentioned above, conventional thermal perturbation theory
exhibits very poor apparent convergence as soon as collective
phenomena such as the plasmon effect are included, which
involve odd powers in the coupling. The self-consistent
entropy and density functionals at two-loop order
also contain the plasmon effect, but together with
higher-order terms that ordinarily would have been discarded
by a strictly perturbative expansion in $g$ because otherwise
the renormalization programme could not have been carried out.
By contrast, the two-loop entropy and density functional is UV finite
and can be evaluated in a nonperturbative manner using
the high-temperature/density approximation of the gluon
and quark propagators, which are finite at leading and
next-to-leading order in HTL perturbation theory.

\subsection{HTL/HDL approximation}

As a first approximation let us consider the two-loop
entropy and density functionals evaluated completely using
HTL (HDL) propagators. 

\subsubsection{Entropy}

In the case of the entropy,
this takes care of all
contributions of order $g^2$, but only part (1/4) of
the plasmon term $\sim g^3$. However, it also contains
infinitely many higher-order terms which despite
being incomplete may help to get rid of the pathological
behaviour of the perturbation series truncated at
low orders in $g$. Among such higher-order contributions
is for instance a $g^4$ contribution reading (for pure glue)
\be\label{S4HTL}
{\cal S}_{HTL}^{(4)}=-N_g{\hat m_D^4 \0 16\pi^2 T}
\(\log{T\0\hat m_D}+1.55\ldots \)
\ee
involving a $g^4\log(1/g)$-term, whose coefficient in pure-glue QCD
is 1/12 of that of the complete perturbative result.\footnote{The 
correct coefficient will be restored by $O(g^4\log(1/g)T^2)$ corrections
to $m_\infty^2$, whereas the constant under the logarithm also
receives three-loop contributions.}
By contrast, simple massive quasiparticle models such as
those used in Refs.~\cite{Peshier:1996ty,Levai:1997yx}
do not have a $g^4\log(1/g)$-term in the entropy at all.


\begin{figure}[htb]
\includegraphics[bb=70 240 540 560,width=6.7cm]{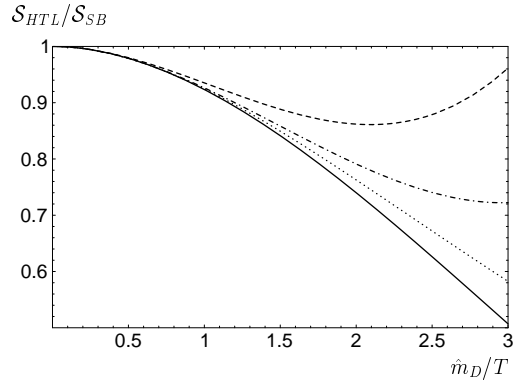}
\caption{HTL entropy per gluonic degree of freedom normalized to
its Stefan-Boltzmann value as a function of the Debye mass 
$\hat m_D(T,\mu)/T$. 
The full line gives the complete numerical
result; the dashed line corresponds
to the perturbative result to order $(\hat m_D/T)^3\sim g^3$. The
dotted line gives the entropy for scalar degrees of freedom with
momentum-independent mass $m=m_\infty=\hat m_D/\sqrt2$; its
perturbative approximant is given by the dash-dotted line.}
\label{figSS0g}
\end{figure}

In Fig.~\ref{figSS0g}, the numerical result for ${\cal S}_{HTL}/{\cal S}_{SB}$
in the case of pure glue is given as function of $\hat m_D/T$, which
is the only independent parameter. The HTL result is given by the
full line and is found to be a monotonicly decreasing function
of $\hat m_D/T$. If this were expanded in powers of $\hat m_D/T$
and truncated beyond $(\hat m_D/T)^3\sim g^3$
(dashed line in Fig.~\ref{figSS0g}), this property would
have been lost at $\hat m_D/(2\pi T)\approx 1/3$, where one might
still expect a sufficiently clear separation of hard and soft
scales which is a prerequisite of the HTL approximation.

The numerical result for ${\cal S}_{HTL}$ is in fact very close
to a simple massive
quasiparticle model with entropy $2N_g {\cal S}_{SB}(m_\infty)$,
represented by the dotted line in Fig.~\ref{figSS0g}.
Remarkably, one has ${\cal S}_{HTL}<2N_g {\cal S}_{SB}(m_\infty)$ even
though the latter contains 30\% less of the plasmon term $\sim g^3$, which
would be expected to work in the opposite direction. 
This is further emphasized by comparison with its perturbative
approximation given by the dash-dotted line.

\begin{figure}[htb]
\includegraphics[bb=70 240 540 560,width=6.7cm]{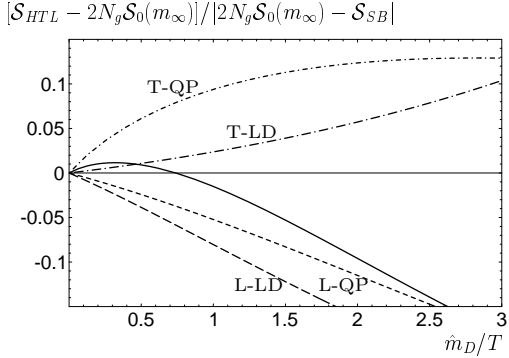}
\begin{picture}(0,0)(0,0)
\put(-140,85){\scriptsize T-QP}
\put(-110,68){\scriptsize T-LD}
\put(-75,10){\scriptsize L-QP}
\put(-107,10){\scriptsize L-LD}
\end{picture}
\caption{Relative deviation of the HTL entropy from that of a gas of massive
bosons with (constant) mass $m_\infty$ (full line) and their
composition from 
transverse (T) and longitudinal (L) quasiparticle (QP) and
Landau damping (LD) contributions.}
\label{figdSdetail}
\end{figure}

The rather small difference to a simple massive quasiparticle
model is further inspected in Fig.~\ref{figdSdetail},
where it is shown that there are large cancellations
between the longitudinal and transverse contributions
to the HTL entropy.

\begin{figure}[htb]
\includegraphics[bb=70 240 540 560,width=6.7cm]{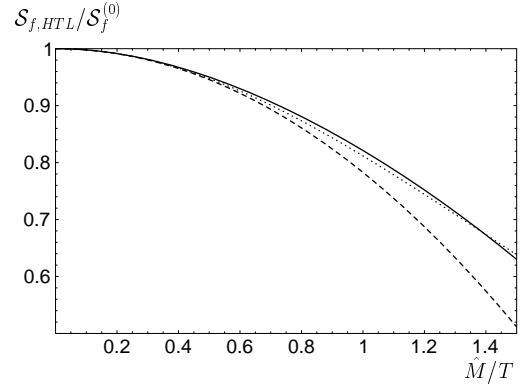}
\caption{HTL entropy per quark degree of freedom at $\mu=0$ normalized
to its free value (solid line), the corresponding perturbative
order-$g^2$ result (dashed line), and the entropy of
a quark with constant mass $M_\infty$ (dotted line).}
\label{figSS0q}
\end{figure}

In Fig.~\ref{figSS0q} the HTL entropy of 1 quark degree of freedom
at zero chemical potential is displayed as a function of
the fermionic plasma frequency $\hat M/T$. Like in the
pure-glue case, this contains also an (incomplete) $g^4\log(1/g)$
contributions that is not present in simpler quasiparticle models:
\be
{\cal S}_{f,HTL}^{(4)}=NN_f{\hat M^4\0\pi^2 T}\( \log{T\0\hat M}+0.22\ldots \)
\ee
${\cal S}_{f,HTL}$ does not contain any contributions to the
plasmon term $\sim g^3$, which entirely come from NLO corrections
to $M_\infty$. There is therefore no big deviation of the
full numerical result (solid line) from the
perturbative one truncated beyond $g^2$ (dashed line).

Compared to the entropy of a simple massive fermionic
quasiparticle model ${\cal S}_{f}^{(0)}(M_\infty)$ (dotted line
in Fig.~\ref{figSS0q}), one finds extremely good numerical agreement,
which again takes place only after adding up all the quasiparticle
($(+)$ and $(-)$) and Landau-damping contributions, however.

\subsubsection{Quark density}

In the case of zero temperature but high chemical potential,
the quantity of interest is the quark density. This does
not contain any plasmon term $\sim g^3$, but rather
$\sim g^4 \log(1/g)$. The HDL approximation does contain
some though not all of this term:
\be
{\cal N}_{HDL}^{(4)}= NN_f{\hat M^4\0\pi^2 \mu}\( \log{\mu\0\hat M}+0.35\ldots \)
\ee
Order-$g^2 \log g$
contributions to the asymptotic masses of the quark and gluon quasiparticles,
still within the framework of the 2-loop $\Phi$-derivable approximation,
are responsible for the
remaining contribution to the coefficient of the $g^4\log g$-term,
while the coefficient under the logarithm also receives 3-loop contributions.

\begin{figure}[htb]
\includegraphics[bb=70 240 540 560,width=6.7cm]{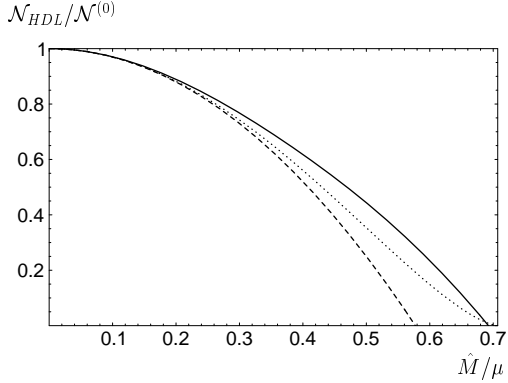}
\caption{HDL quark density at $T=0$ normalized
to its free value (solid line), the corresponding perturbative
order-$g^2$ result (dashed line), and the free quark density of
a quark with constant mass $M_\infty$ (dotted line).}
\label{figNN0}
\end{figure}

In Fig.~\ref{figNN0} the numerical result for ${\cal N}_{HDL}$ is given
as a function of $\hat M/\mu$ (full line) and compared to that of
a simple quasiparticle model
\be\label{NT0}
{\cal N}_0(M_\infty)\Big|_{T=0}={1\03\pi^2}(\mu^2-M_\infty^2)^{3\02} 
\theta(\mu-M_\infty)
\ee
as well as to a perturbative approximation truncated beyond $(\hat M/\mu)^2\sim
g^2$. Remarkably, the full HDL result drops to zero at almost the
same value as the simple quasiparticle model. However, the former
becomes negative thereafter, showing that the HDL approximation
is breaking down at $M_\infty/\mu > 1$ at the latest.

\subsection{Next-to-leading approximations}

The plasmon term $\sim g^3$ at high temperatures
$T\gg \hat m_D$ becomes complete only after inclusion
of the next-to-leading correction to the asymptotic thermal
masses $m_\infty$ and $M_\infty$. These are determined
in standard HTL perturbation theory through
\be
\begin{array}{l}
\delta m_\infty^2(k)=\Re \delta\Pi_T(\omega=k) \\
=\Re(\begin{picture}(0,0)(0,0)
\put(25,0){\small +}
\put(56,0){\small +}
\put(104,0){\small +}
\put(157,0){$|_{\omega=k}$}
\end{picture}
\!\!\includegraphics[bb=145 430 500 475,width=5.5cm]{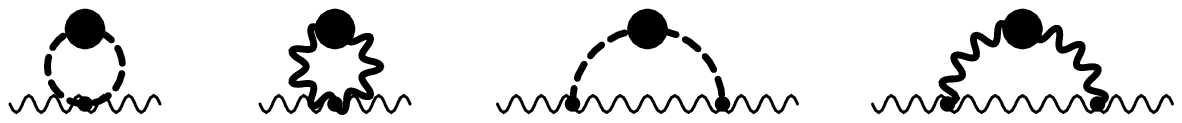})
\end{array}
\ee
where thick dashed and wiggly lines with a blob represent
HTL propagators for longitudinal and transverse polarizations, respectively.
Similarly,
\be
\begin{array}{l}
{1\02k}\delta M_\infty^2(k)=\delta\Sigma_+(\omega=k) \nonumber\\
=\Re(\begin{picture}(0,0)(0,0)
\put(42,0){\small +}
\end{picture}
\includegraphics[bb=75 430 285 475,width=3.2cm]{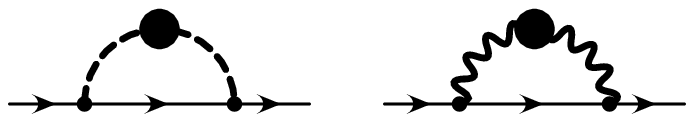})
|_{\omega=k}\;.
\end{array}
\ee

The explicit proof that these contributions indeed restore the
correct plasmon term is given in Ref.~\cite{Blaizot:2000fc}.

These corrections to the asymptotic thermal masses are, in contrast
to the latter, nontrivial functions of the momentum, which can
be evaluated only numerically. However, as far as the generation
of the plasmon term is concerned, these functions contribute
in the averaged form
\be
\bar\delta m_\infty^2={\int dk\,k\,n_{\rm BE}'(k) \Re \delta\Pi_T(\omega=k) 
\0 \int dk\,k\,n_{\rm BE}'(k)}
\ee
(cp.~eq.~\ref{SNBF2}) and similarly
\be
\bar\delta M_\infty^2={\int dk\,k\,n_{\rm FD}'(k) \Re 
2k \delta\Sigma_+(\omega=k) 
\0 \int dk\,k\,n_{\rm FD}'(k)}\;.
\ee
These averaged asymptotic thermal masses turn out to be given
by the remarkably simple expressions
\bea
\label{deltamas}
\bar\delta m_\infty^2=-{1\02\pi}g^2NT\hat m_D,&&\\
\label{deltaMas}
\bar\delta M_\infty^2=-{1\02\pi}g^2C_fT\hat m_D,&&\!\!\!\! C_f=N_g/(2N).\;
\eea

These corrections only pertain to the hard excitations;
corrections to the various thermal masses of soft excitations
are known to differ substantially from (\ref{deltamas})
or (\ref{deltaMas}). For instance, the gluonic
plasma frequency at $k=0$ reads \cite{Schulz:1994gf}
$\delta m^2_{pl.}/\hat m^2_{pl.}\approx
-0.18\sqrt{N}g$, which is only about a third of 
$\bar\delta m_\infty^2/m_\infty^2$; the NLO correction to the
nonabelian Debye mass on the other hand is even positive
for small coupling and moreover logarithmically enhanced
\cite{Rebhan:1993az}
$\delta m^2_D/\hat m^2_D = +\sqrt{3N}/(2\pi) \times g \log(1/g)$.

For an estimate of the effects of a proper incorporation of the
next-to-leading order corrections we have therefore proposed
to include the latter only for hard excitations and to
define our next-to-leading approximation (for gluons) through
\be
{\cal S}_{NLA}={\cal S}_{HTL}\Big|_{\rm soft}+
{\cal S}_{\bar m_\infty^2}\Big|_{\rm hard},
\ee
where $\bar m_\infty^2$ includes (\ref{deltamas}).
To separate soft ($k\sim \hat m_D$) and hard ($k \sim 2\pi T$) momentum
scales, we introduce the intermediate scale
$\Lambda=\sqrt{2\pi T\hat m_D c_\Lambda}$
and consider a variation of $c_\Lambda=\2\ldots 2$ as part of
our theoretical uncertainty.

Another crucial issue concerns the definition of the corrected
asymptotic mass $\bar m_\infty$. For the range of coupling constants
of interest ($g> 1$), the correction $ |\bar\delta m_\infty^2|$
is greater than the LO value $m_\infty^2$, leading to tachyonic
masses if included in a strictly perturbative manner.

However, this problem is not at all specific to QCD. In the
simple $g^2\varphi^4$ model, one-loop resummed perturbation theory
gives 
\be\label{mstrpert}
m^2=g^2T^2(1-{3\0\pi}g)
\ee 
which also turns tachyonic
for $g>1$. On the other hand, the solution of the corresponding
simple scalar gap equation is a monotonic function in $g$, and
it turns out that the first two terms in a $(m/T)$-expansion of
this gap equation,
\be\label{mtruncgap}
m^2=g^2T^2-{3\0\pi}g^2Tm
\ee
which is perturbatively equivalent to (\ref{mstrpert}) has
a solution that is extremely close to that of the full gap
equation (for $\overline{\mbox{MS}}$ renormalization scales $\bar\mu \approx 2\pi T$).

In QCD, the (non-local) gap equations are way too complicated to
be attacked directly. We instead consider perturbatively equivalent
expressions for the corrected $\bar m_\infty$ which are monotonic
functions in $g$. Besides the solution to a quadratic equation
analogous to (\ref{mtruncgap}) we have tried the
simplest Pad\'e approximant $m^2=g^2T^2/(1+{3\0\pi}g)$, which
also gives a greatly improved approximation to the solution of
scalar gap equations. In QCD, our final results do not depend
too much on whether we use the Pad\'e approximant
\cite{Blaizot:1999ip,Blaizot:1999ap}
or a quadratic gap equation
\cite{Blaizot:2000fc}.

The main uncertainty rather comes from the choice of the
renormalization scale which determines the magnitude of
the strong coupling constant when this is taken as determined
by the renormalization group equation (2-loop in the following).

\begin{figure}[tb]
\includegraphics[bb=70 240 540 560,width=6.7cm]{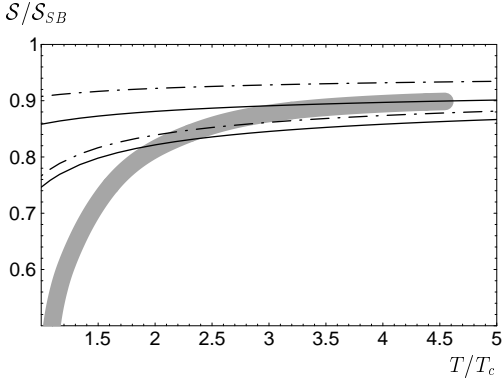}
\caption{Comparison of the lattice data for
the entropy of pure-glue SU(3) gauge theory of Ref.~\cite{Boyd:1996bx}
(gray band) with the range of ${\cal S}_{HTL}$ (solid lines)
and ${\cal S}_{NLA}$ (dash-dotted lines) for $\bar\mu=
\pi T\ldots 4\pi T$ and $c_\Lambda= 1/2 \ldots 2$.
}
\label{figSg}
\end{figure}

In Fig.~\ref{figSg}, the numerical results for the HTL entropy
and the NLA one are given as a function of $T/T_c$ with $T_c$
chosen as $T_c=1.14\Lambda_{\overline{\mathrm MS}}$. The full lines
show the range of results for
${\cal S}_{HTL}$ when the renormalization scale $\bar\mu$
is varied from $\pi T$ to $4\pi T$; the dash-dotted lines mark
the corresponding results for ${\cal S}_{NLA}$ with the
additional variation of $c_\Lambda$ from $1/2$ to 2. The
dark-gray band are lattice data from Ref.~\cite{Boyd:1996bx};
the more recent results from Ref.~\cite{Okamoto:1999hi} are consistent
with the former within error bars and centered around the upper
boundary of the gray band for $T\approx 3T_c$ and somewhat flatter
around $2T_c$.
Evidently, there is very good agreement for $T>2.5T_c$.

\begin{figure}[tb]
\includegraphics[bb=70 240 540 560,width=6.7cm]{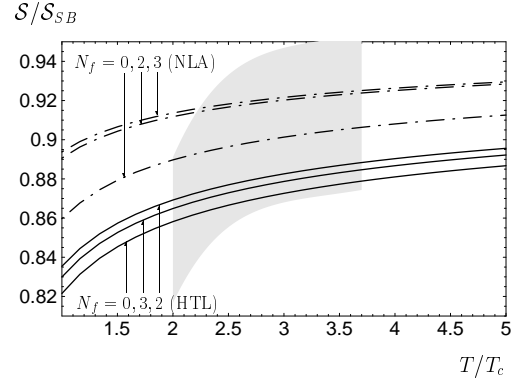}
\caption{Comparison of the HTL entropy (solid lines) and the NLA
estimate (dash-dotted lines) for $N_f=0,2,3$ with
the estimated continuum extrapolation of $N_f=2$ lattice data
of Ref.~\cite{Karsch:1999vy}.}
\label{figSNf023f}
\end{figure}

In Fig.~\ref{figSNf023f}, $N_f$ massless quarks are included and
compared with a recent estimate \cite{Karsch:1999vy} of the continuum limit of
lattice results for $N_f=2$ (gray band), 
but now with ${\cal S}_{HTL}$ and ${\cal S}_{NLA}$ 
evaluated for the central choice
of $\bar\mu=2\pi T$ and $c_\Lambda=1$ (with unchanged
$T_c/\Lambda_{\overline{\mathrm MS}}$). When $N_f$ is increased,
there are competing effects of larger thermal masses versus
slower running of $\alpha_s$, which result into a rather
weak dependence of our results on $N_f$ as a function
of $T/\Lambda_{\overline{\mathrm MS}}$ as it is in Fig.~\ref{figSNf023f}.

{}From the above results for the entropy density, one can
recover the thermodynamic pressure by simple integration,
\be
P(T)-P(T_0)=\int_{T_0}^T dT' S(T').
\ee
The integration constant $P(T_0)$, however, is a
strictly nonperturbative input. It cannot be fixed by
requiring $P(T=0)=0$, as this is in the confinement regime.
It is also not sufficient to know that $\lim_{T\to\infty}P=P_{\rm free}$
by asymptotic freedom. In fact, the undetermined integration constant
in $P(T)/P_{\rm free}(T)$ when expressed as a function
of $\alpha_s(T)$ corresponds to a term \cite{Blaizot:1999ap}
\be
C\exp\(-{1\0\alpha_s}
[4\beta_0^{-1}+O(\alpha_s)]\)
\ee
which vanishes for $\alpha_s\to 0$ with all derivatives
and thus is not fixed by any order of perturbation theory.
It is, in essence, the nonperturbative bag constant, which
can be added on to standard perturbative results, too. However,
in $P(T)/P_{\rm free}(T)$ this term becomes rapidly unimportant
as the temperature is increased, as it decays like $T^{-4}$.

\begin{figure}[tb]
\includegraphics[bb=20 210 540 540,width=7.2cm]{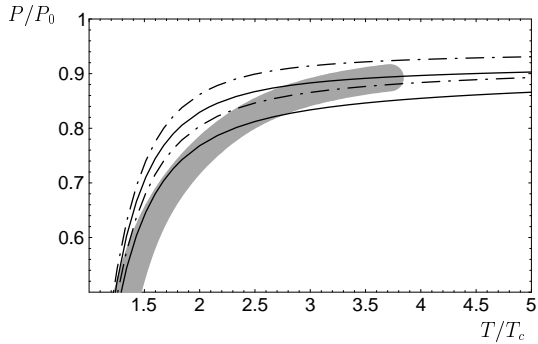}
\caption{Comparison of our results for the pressure of QCD with
$N_f=2$ with the extrapolated lattice data of Ref.~\cite{Karsch:1999vy}.}
\label{P2}
\end{figure}

In Fig.~\ref{P2} which shows the HTL and NLA pressure
in the case of $N_f=2$, we have fixed this integration constant by
choosing $P(T_c)=0$ as $T=T_c$ is beyond the range of applicability
of any form of perturbation theory anyway. 
As with the entropy, there is good agreement with lattice data
for $T>2.5T_c$, and this agreement does not depend on the
precise value of $P(T_c)$, which enters only at the percent level
at such temperatures.

\section{Discussion and outlook}

Starting from the self-consistent two-loop approximations to
entropy and density, we have been able to give a quasiparticle
description of the thermodynamics of QCD that agrees well
with available lattice data for $T>2.5T_c$, and which
incorporate more of the perturbatively accessible details
of gluonic and fermionic quasiparticles than previously
proposed quasiparticle models 
\cite{Peshier:1996ty,Levai:1997yx,Schneider:2001nf}
involving simply free massive
scalars and fermions. 
In contrast to the latter, however, we refrained from introducing
phenomenological functions to fit the data, the only
input being $T_c/\Lambda_{\overline{\mathrm MS}}$ as gleaned from the lattice.
Put conservatively, we obtained a lower bound on $T/T_c$
for which HTL quasiparticles may be an appropriate description
of hot QCD. This lower bound is however extremely low when
compared with what is needed for conventional thermal perturbation
theory to avoid an obvious breakdown, raising the hope that
the latter just needs further resummations to become applicable
at temperatures of physical interest in QCD.

There exist in fact alternative proposals for such resummations.
One particularly interesting approach is an extension of
so-called screened perturbation theory \cite{Karsch:1997gj,Andersen:2000yj},
where thermal masses are introduced as variational parameters,
to nonabelian gauge theories, using the HTL effective action
as gauge invariant thermal mass term \cite{Andersen:1999}.
It should be noted, however, that the latter is introduced
as a mere technical device, whereas in the entropy-based approach
the emphasis is on a description in terms of HTL quasiparticles
and their perturbatively accessible modifications at NLO.
Indeed, in the two-loop entropy it turned out that at hard
momentum scales only the region close to the light-cone
matters, where the HTL approximations remain accurate even
at hard momenta. By contrast, in HTL-screened perturbation theory,
the HTL's contribute throughout all of phase space. Moreover,
the thus modified UV structure entails new UV problems (together
with new scheme dependences).

To avoid these UV problems, Peshier \cite{Peshier:2000hx} has
proposed a somewhat contrived way of HTL resummation directly on
the thermodynamic pressure, which appears to come very close
to what we have obtained in the HTL approximation
in the entropy. Peshier's pressure is perturbatively equivalent
to ours in the HTL approximation through at least order $g^3$,
though not obviously identical to it.

Beyond the HTL approximation, it is not clear how the
formulae of Ref.~\cite{Peshier:2000hx} could be extended.
In the present HTL-based quasiparticle approach, the aim is to
describe these quasiparticles in as full detail as
HTL perturbation theory permits. In this respect we have
so far taken only a first step by considering averaged
asymptotic thermal masses. We intend to
study the full momentum dependence of the asymptotic thermal
masses and how this enters into the thermodynamic quantities
in a future work.

Another extension of the present work will be to consider
in more detail the dependence on chemical potential, for example
by integrating entropy and density for all $T$ and $\mu$ in
the deconfined phase, similarly to what has been
carried out in simple quasiparticle models \cite{Peshier:1999ww}. 
Using the NLA, this will include a complete plasmon term,
and requires only
one integration constant, $P(T_0)|_{\mu=0}$, that can be
taken from the lattice.

A first (infinitesimal) step towards finite $\mu$ can be
studied on the lattice in the form of quark number
susceptibilities $\6{\cal N}/\6 \mu |_{\mu=0}$ at high $T$
\cite{Gavai:2001fr,Gavai:2001BI}. Here conventional thermal
perturbation theory is as ill-behaved as in the case
of the pressure, whereas our approach can be applied
in a straightforward manner, as will be presented in
a forthcoming paper \cite{Blaizot:2001chi}.

\section*{Acknowledgements}

The work presented here has been carried out in a most
enjoyable collaboration
with Jean-Paul Blaizot and Edmond Iancu.
I would also like to thank the organizers of
``Statistical QCD'', Frithjof Karsch and Helmut Satz,
for their hospitality at this stimulating meeting.




\bibliography{tft,tftpr,qft,ar,books}      
\bibliographystyle{elsart-numwot}

\end{document}